\documentclass[onecolumn]{aa}

\usepackage{xspace}
\usepackage{graphicx}
\usepackage{float}

\usepackage{amssymb}
\usepackage{txfonts}
\usepackage{multicol}
\usepackage{subfig}

\usepackage{hyperref}
\usepackage{natbib}
\bibpunct{(}{)}{;}{a}{}{,} 

\usepackage{afterpage}
\usepackage{textcomp}
\usepackage{xcolor}
\usepackage{soul,xcolor}
\usepackage{xtab}

\usepackage{lipsum}
\newenvironment{Figure}
  {\par\medskip\noindent\minipage{\linewidth}}
  {\endminipage\par\medskip}

\usepackage{placeins}

\newcommand{\cc}{\ensuremath{{\rm cm}^{-3}}\xspace}
\newcommand{\amm}{\ensuremath{{\rm NH}_3}\xspace}
\newcommand{\tex}{\ensuremath{T_{\rm ex}}\xspace}
\newcommand{\tk}{\ensuremath{T_{\rm K}}\xspace}
\newcommand{\sig}{\ensuremath{\sigma_{\rm{v}}}\xspace}
\newcommand{\vel}{\ensuremath{\rm{v}_{LSR}}\xspace}
\newcommand{\htw}{\ensuremath{\rm H_2}\xspace}
\newcommand{\msun}{\ensuremath{M_\odot}\xspace}
\newcommand{\msunyr}{\ensuremath{M_\odot\,{\rm yr}^{-1}}\xspace}

\newcommand{\kms}{\ensuremath{{\rm km\,s}^{-1}}\xspace}

\newcommand{\cv}{\ensuremath{C_{\rm v}}\xspace}

\begin{document}

  \title{Infall of material onto the filaments in Barnard 5}
  
   \author{
        Spandan Choudhury
          \inst{1,2}
          \and
          Jaime E. Pineda \inst{1}
          \and
          Paola Caselli \inst{1}
          \and
          Michael Chun-Yuan Chen \inst{3}
          \and
          Stella S. R. Offner \inst{4}
          \and
          Maria Teresa Valdivia-Mena\inst{1}
          }

    \institute{
        Max-Planck-Institut f\"ur extraterrestrische Physik, Giessenbachstrasse 1, D-85748 Garching, Germany
        \and
        Korea Astronomy and Space Science Institute, 776 Daedeok-daero Yuseong-gu, Daejeon 34055, Republic of Korea \\
        \email{spandan@kasi.re.kr}
        \and
        The Department of Physics, Engineering Physics \& Astronomy, Queen's University, Stirling Hall, 64 Bader Lane, Kingston, ON K7L 3N6, Canada
        \and
        Department of Astronomy, The University of Texas at Austin, Austin, TX 78712, USA
        }
    \date{}

 
  \abstract
  {}
  {We aim to study the structure and kinematics of the two filaments inside the subsonic core Barnard 5 in Perseus using high-resolution ($\approx$ 2400 au) \amm data and a multi-component fit analysis.}
  {We used observations of \amm (1,1) and (2,2) inversion transitions using the Very Large Array (VLA) and the Green Bank Telescope (GBT). We smoothed the data to a beam of 8$\arcsec$ to reliably fit multiple velocity components towards the two filamentary structures identified in B5.}
  {Along with the core and cloud components, which dominate the flux in the line of sight, we detected two components towards the two filaments showing signs of infall. We also detected two additional components that can possibly trace new material falling into the subsonic core of B5.}
  {Following comparison with previous simulations of filament formation scenarios in planar geometry, we conclude that either the formation of the B5 filaments is likely to be rather cylindrically symmetrical or the filaments are magnetically supported.
  We also estimate infall rates of $1.6\times10^{-4}\,$\msunyr and $1.8\times10^{-4}\,$\msunyr (upper limits) for the material being accreted onto the two filaments. At these rates, the filament masses can change significantly during the core lifetime.
  We also estimate an upper limit of $3.5\times10^{-5}\,$\msunyr for the rate of possible infall onto the core itself. Accretion of new material onto cores indicates the need for a significant update to current core evolution models, where cores are assumed to evolve in isolation.
  }

   \keywords{ ISM: kinematics and dynamics -- ISM: individual objects (B5, Perseus) -- ISM: molecules -- star: formation}

   \maketitle
%

\section{Introduction}

Stars form in cold dense cores embedded in molecular clouds. These cores are characterised by higher densities and lower temperatures, compared to the parental cloud
\citep{myers_1983_SF_core,myers-benson_1983_dense_core,caselli_2002_dense_core}. Studies with high-density ($n(\htw) > 10^4\,$\cc) tracers have revealed subsonic levels of turbulence inside cores \citep{coh_core_barranco_goodman_1998,roso_2008_amm_dens_core}, in contrast to the supersonic linewidths observed in the ambient cloud \citep[traced by lower density tracers, like CO;][]{larson_81}.

A useful tracer for observing dense cores in molecular clouds is \amm. Even though it has a relatively low critical density (a few times $\rm 10^3\,$\cc), the hyperfine structure of the \amm\ inversion transitions allows for the individual hyperfines to remain optically thin even at high column densities \citep{caselli_2017_amm}. Fitting the multiple hyperfines of \amm\ simultaneously also gives a very precise constraint on the kinematical information. Moreover, if the \amm\ (2,2) line is reliably detected along with the (1,1), it allows for a direct measurement of the gas temperature \citep[e.g. ][]{GASDR1}.

Previous observations have shown filaments to be widely present in molecular clouds in star-forming regions and connected to subsonic dense cores \citep{goldsmith_2008_fila, andre_2010_fila, poly_2013_fila}. These filamentary structures have been extensively observed using $\emph{Herschel}$ \citep{wt_2010_fila, ppvi_andre_2014_fila, arzou_2019_fila} and studied with respect to a range of physical properties, including density and temperature. Molecular line observations of filaments have revealed complex kinematical structures \citep{kirk_2007_rel_vel, henshaw_2013_fila, hacar_2013_fila}. 
Using molecular lines, filaments have been observed to have much smaller lengths \citep{hacar_2017_small_fila, monsch_2018_small_fila, suri_2019_small_fila} than the filaments identified with continuum observations. These filaments have been observed with lengths ranging from 0.01 pc to 0.3 pc; therefore, they appear to be scale-free.

\citet{chen_2020_fila_form} suggested the use of a dimensionless parameter, \cv, to distinguish between two scenarios resulting in the formation of filaments, namely, turbulent-driven and gravity-dominated compression. 
This parameter, which depends on the velocity gradient orthogonal to the major axis of the filament and the mass per unit length of the filament,
has been used to differentiate between the formation scenarios of filaments in observations of star-forming regions \citep[e.g.][]{zhang_2020_cv_used, gong_2021_cv_use, hsieh_2021_cv_use}. 
However, in these studies, a single value is calculated representing the entire filament and no distribution along the filament is reported. Analyses of velocity gradients along the filament could highlight the robustness of this parameter, as well as to study the overall structure of the filaments.

We study the subsonic core Barnard 5 (hereafter B5) in the Perseus molecular cloud with combined Very Large Array (VLA) and Green Bank Telescope (GBT) observations using \amm\ (1,1) and (2,2) transitions. B5 is located at a distance of $302 \pm 21\, \rm{pc}$ \citep{zucker2018_pers_dist}, and hosts a young stellar object (YSO), the Class-I protostar B5-IRS1 \citep{fuller1991_B5IRS1}. Using \amm\ hyperfine transitions, \citet{pineda2010} revealed the turbulence inside the core to be subsonic and nearly constant, with a sharp transition to coherence at the core boundary. Further interferometric observations with high spatial resolution (6$''$,  $\approx1800$\,au) revealed two filaments embedded in the subsonic core \citep{pineda2011_B5_fila} and three gravitationally bound condensations \citep{pineda2015_B5conden, anika_2021}. 
These filaments are similar in length to those presented in \citet{suri_2019_small_fila}. However, it is not yet well understood how such subsonic filaments within dense cores relate to their large-scale counterparts.
In this project, we use this high-resolution data, smoothed to a slightly larger beam (8$''$), to study the detailed velocity structure within B5 and the embedded filaments with multi-component analysis.


\section{Ammonia maps}
\label{sec_amm_data}

We used single-dish observations of B5 with the GBT combined with interferometric observations with the VLA of \amm\ inversion transitions (1,1) and (2,2). The GBT observations were performed under project 08C-088 with on-the-fly (OTF) mapping \citep{mangum2007_otf} in frequency-switching mode. The data have a spectral resolution of about 3 kHz (or 0.04 \kms). More details of the data processing can be found in \citet{pineda2010}, where the data were first published. The interferometric data were obtained with the VLA in the D-array configuration on 16-17 October 2011 and in the DnC-array configuration on 13-14 January 2012  (project 11B-101). The spectral resolution of the data was 0.049 \kms. A detailed description of the data can be found in \citet{pineda2015_B5conden}. The resolution of the combined data is 6$''$.

\begin{figure*}[!ht]  
\centering
\includegraphics[width=0.76\textwidth]{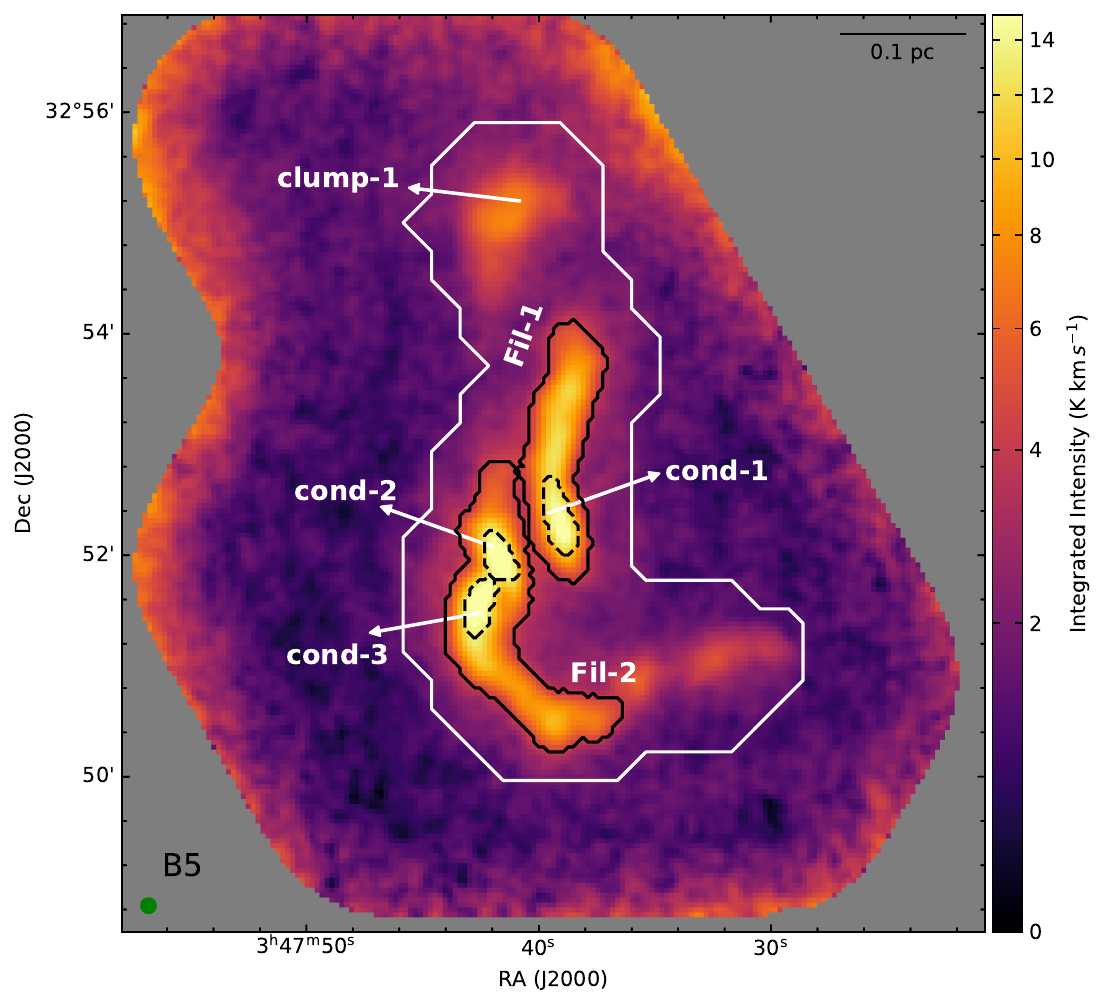}
  \caption{Integrated intensity of \amm\ (1,1) in Barnard 5. The black-solid and black-dashed contours show the two filaments and the three condensations in B5, respectively \citep[contours and nomenclature adapted from][]{anika_2021}. 
  The beam size and scale-bar are shown in the bottom-left and the top-right corners, respectively. The white-solid contour shows the core boundary calculated using single-dish observations by \citet{pineda2010}.} 
     \label{fig_mom0} 
\end{figure*}

We smoothed the data to a beam of 8$''$ to increase the sensitivity, which enables us to identify and fit multiple components towards the filaments.
To avoid oversampling, we fixed the pixel-to-beam ratio to 3. In this work, we follow the nomenclature of \citet{pineda2015_B5conden} and \citet{anika_2021} regarding the different regions in B5 clump 1, filaments 1 and 2, and condensations 1, 2, and 3). Figure \ref{fig_mom0} shows the integrated intensity map of \amm\ (1,1) in B5, along with the locations of the two filaments and the three condensations inside the subsonic core.

\section{Analysis}
\label{sec_analy}

\subsection{Line fitting}
\label{sec_line_fit}

We used the \verb+cold_ammonia+ model in the python package \verb+pyspeckit+ \citep{pyspeckit, pysp_2022} to fit the \amm\ (1,1) and (2,2) data \footnote{While fitting two (or three) components to the data, we add a two-component (or three-component) model as a new model in the fitter, and use this new model for the fit.}. This model is appropriate for regions with temperatures $<40\, K$ \citep[see][]{GASDR1}. The model fits the \amm (1,1) and (2,2) spectra simultaneously with kinetic and excitation temperatures (\tk\ and \tex), \amm\ column density (N(\amm)), velocity dispersion (\sig), centroid velocity (\vel), and the ortho-\amm\ fraction of the total \amm column density ($\rm f_{ortho}$) as parameters.

We followed the Bayesian approach detailed in \citet{vlas_2020_bayesian} to fit multiple components and selected the optimum number of components in the model. In classical line fitting procedures, the best-fit model is obtained by minimising the $\chi^2$ parameter:
\begin{equation}
    \chi^2 (\theta) = \sum_{i} \frac{(I_i - \mathcal{M}_i( \theta))^2}{\sigma^2} ~,
\end{equation}
where $I_i$ and $\mathcal{M}_i(\theta)$ are the observed and  model intensity in channel $i,$ respectively, and $\sigma$ is the noise in the data, while $\theta$ represents the parameters (see above) that define the model. $\chi^2$ is related to the likelihood function, $\mathcal{L(\theta)}$, as follows:
\begin{equation}
    \textrm{ln}\ \mathcal{L(\theta)} = \frac{\chi^2}{2} + const.
\end{equation}
In the model selection employed in our work, the Bayes factor $K_b^a$ gives the estimate of whether model $\mathcal{M}_a$ is favoured over model $\mathcal{M}_b$ :
\begin{equation}
\label{eq_k_factor}
    K_b^a = \frac{P(\mathcal{M}_b)Z_a}{P(\mathcal{M}_a)Z_b} ~,
\end{equation}
where $P(\mathcal{M}_l)$s are the posterior probability densities and $Z_l$s are the likelihood integrals over the parameter space. The likelihood integral, $Z$, is given by :
\begin{equation}
    Z = \int_\theta p(\theta) \mathcal{L(\theta)} d\theta ~,
\end{equation}
where $p(\theta)$ are the prior probability densities of the parameters represented by $\theta$.
The posterior probability densities can be calculated using the likelihood function as :
\begin{equation}
    P(\theta) \propto p(\theta) \times \mathcal{L(\theta)} ~.
\end{equation}
We adopted a threshold of $\rm{\ln K^a_b=5}$ to indicate whether model $\mathcal{M}_a$ is preferred over $\mathcal{M}_b$ \citep[following][]{vlas_2020_bayesian}. Figure \ref{fig_k-factors} shows the detailed maps of the factors $K^1_0$, $K^2_1$, and $K^3_2$.

The initial priors were estimated to accommodate the ranges of these parameters in B5 known from previous works with single-component fits \citep{pineda2011_B5_fila, pineda_2021_B5, anika_2021}. The ranges of the priors were suitably increased if they were found to be inadequate for the multi-component fits.
As the priors in the model, we use uniformly distributed ranges of $6-15\, $K for \tk, $4-12\, $K for \tex, $10^{13.2} - 10^{15}\, \rm cm^{-2}$ for $ N_{\amm}$, $0.04 - 0.3\,$\kms\ for \sig\ and $9-11\,$\kms\ for \vel.
Since we only have detections of the ortho-\amm (1,1) and (2,2) lines, we cannot fit for $\rm f_{ortho}$. Therefore, for each fit, the ortho-ammonia fraction was fixed at 0.5 (for an assumed ortho-para ratio of unity). Additionally, in order to reduce the likelihood of spurious components, the minimum separation between two components was required to be 0.1 \kms\ (approx. twice the spectral resolution); while the maximum separation between two components was set at 2 \kms\ \citep[more than double the maximum range in velocity across B5 from single component fit results, see][]{pineda2010, anika_2021} to reduce the parameter space to be covered in the likelihood calculations (this was achieved by suitable priors for the additional components).

\begin{figure*}[!ht]  
\centering
\includegraphics[width=\textwidth]{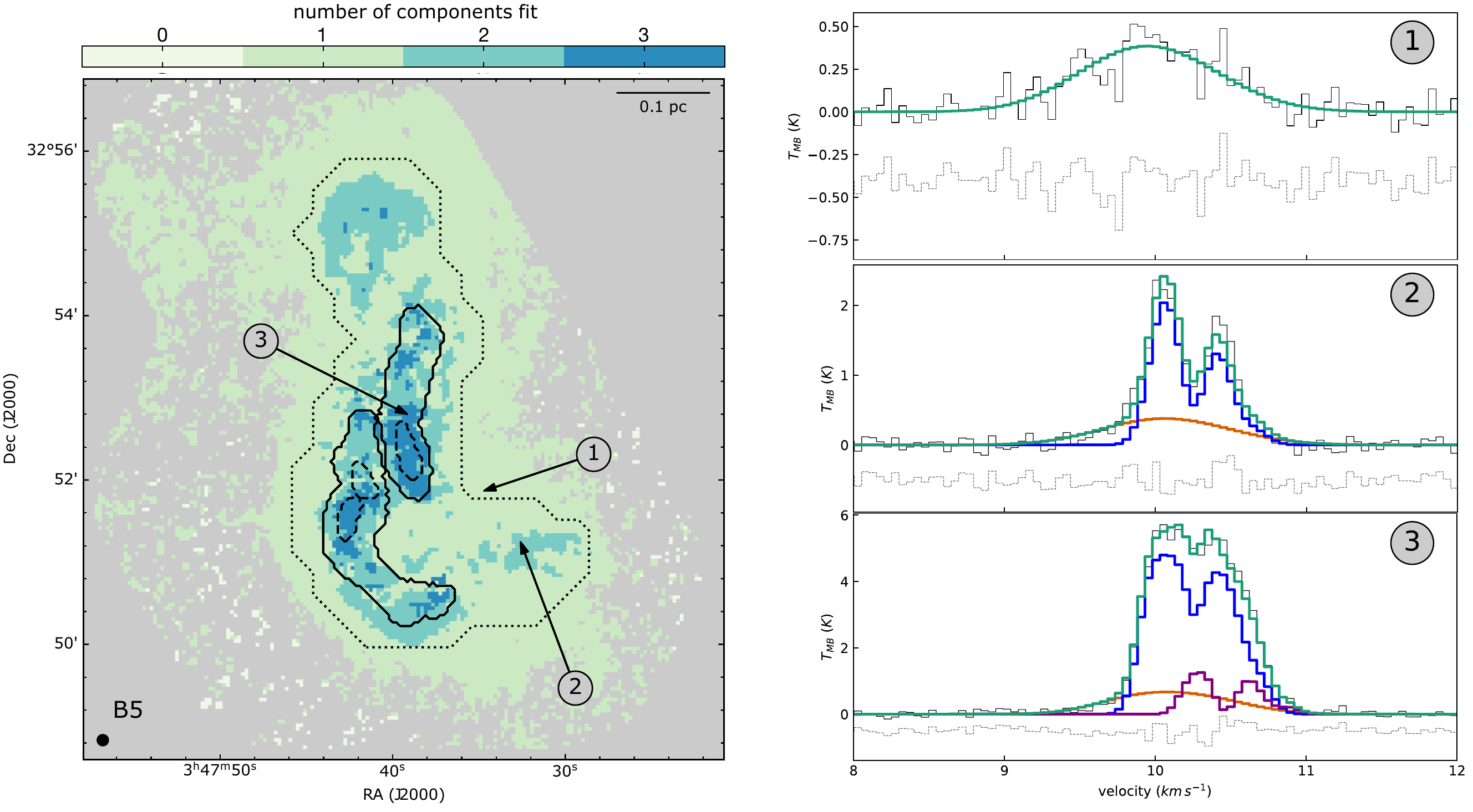}
  \caption{Number of components detected towards different parts of B5, as decided by the Bayesian approach described in Section \ref{sec_line_fit} (left). These components are then subdivided into 5 separate components according to their velocities (see Section \ref{sec_comp_sort}).
  The black-solid and black-dashed contours show the two filaments and the three condensations in B5, respectively \citep[contours adapted from][]{anika_2021}. The black-dotted contour shows the boundary of the coherent core in \citet{pineda2010}.
  The beam size and scale-bar are shown in the bottom-left and the top-right corners, respectively. Examples of \amm (1,1) spectra from regions with one-, two-, and three-component fits (right, with the positions of the spectra indicated in the left panel). 
  Only a small range in velocity is shown here to clearly highlight the individual components, which are illustrated in blue, red, and purple.
  In each panel, the fitted model is shown in green, while the residuals are shown in grey. Note: there are multiple, closely-spaced hyperfines in the velocity range shown here. The strongest two of these hyperfines can be seen in the narrow component (blue and purple in panels 2 and 3). 
  However, these hyperfines are blended together in the broad component (green in panel 1, red in panels 2 and 3), and therefore the hyperfine structure is not visible.}
     \label{fig_num_of_comp}
\end{figure*}

After the initial runs, we identified that all pixels with a good 1-, 2- and 3-component fits are within a signal-to-noise ratio (S/N) of 4, 7, and 10, respectively\footnote{From visual inspection of the fits at different pixels, we found these S/N cuts to be good indication of pixels with reliable fits with the respective number of components.}, in the peak intensity of the \amm (1,1) line. We therefore apply these thresholds in the parameter maps presented here. 
We attempted to fit up to four components to the data, and found that three was the maximum number of components in the best-fit model for regions larger than a beam. Figure \ref{fig_num_of_comp} shows the number of components in the best-fit model in different parts of B5. Throughout the core, we are able to fit at least one component (see Fig. \ref{fig_num_of_comp}). Additionally, towards the two filaments and clump 1, we could fit two components and in some regions inside the filaments, we find that a three-component fit is the optimal model.

\subsection{Component assignment}
\label{sec_comp_sort}

\begin{figure*}[!ht]  
\centering
\includegraphics[width=0.50\textwidth]{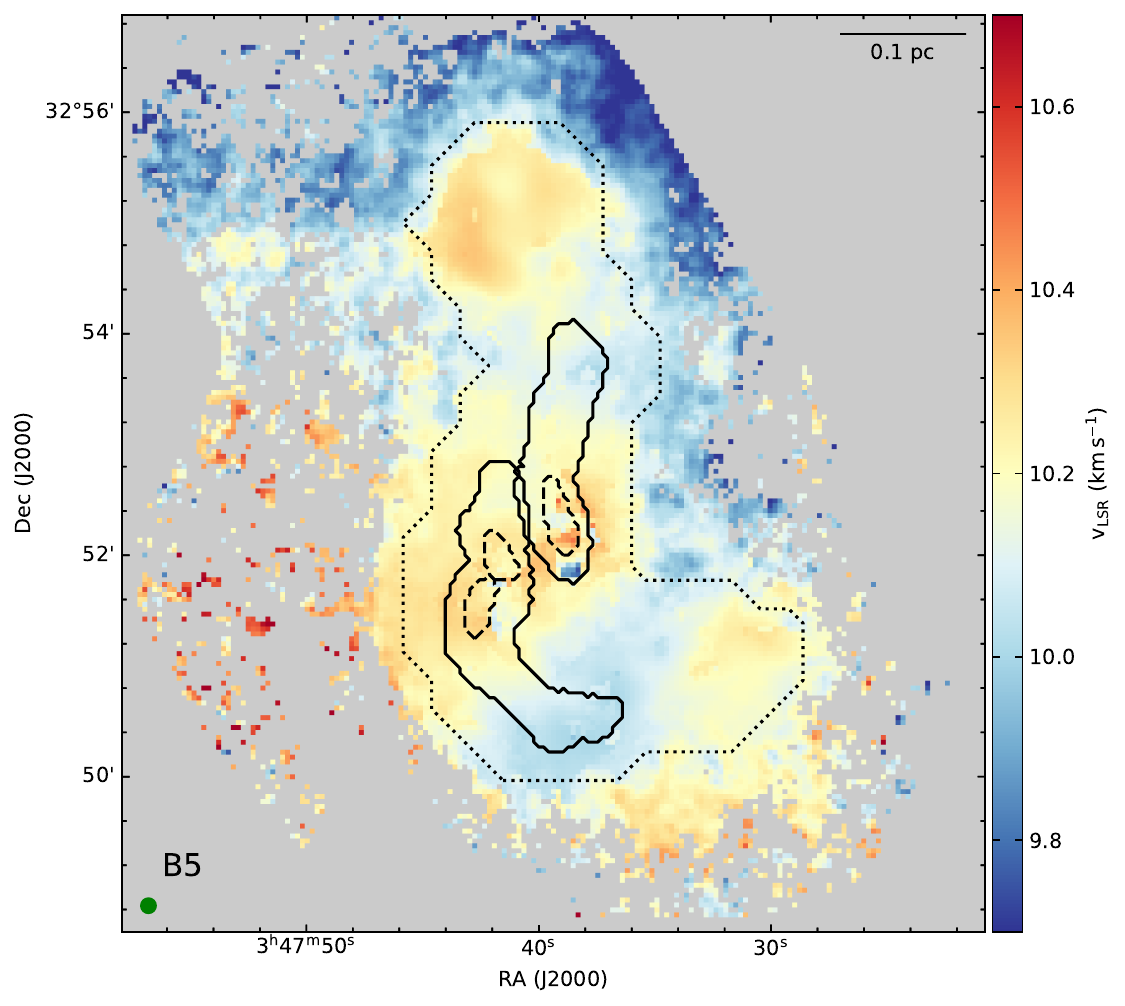}
\includegraphics[width=0.48\textwidth]{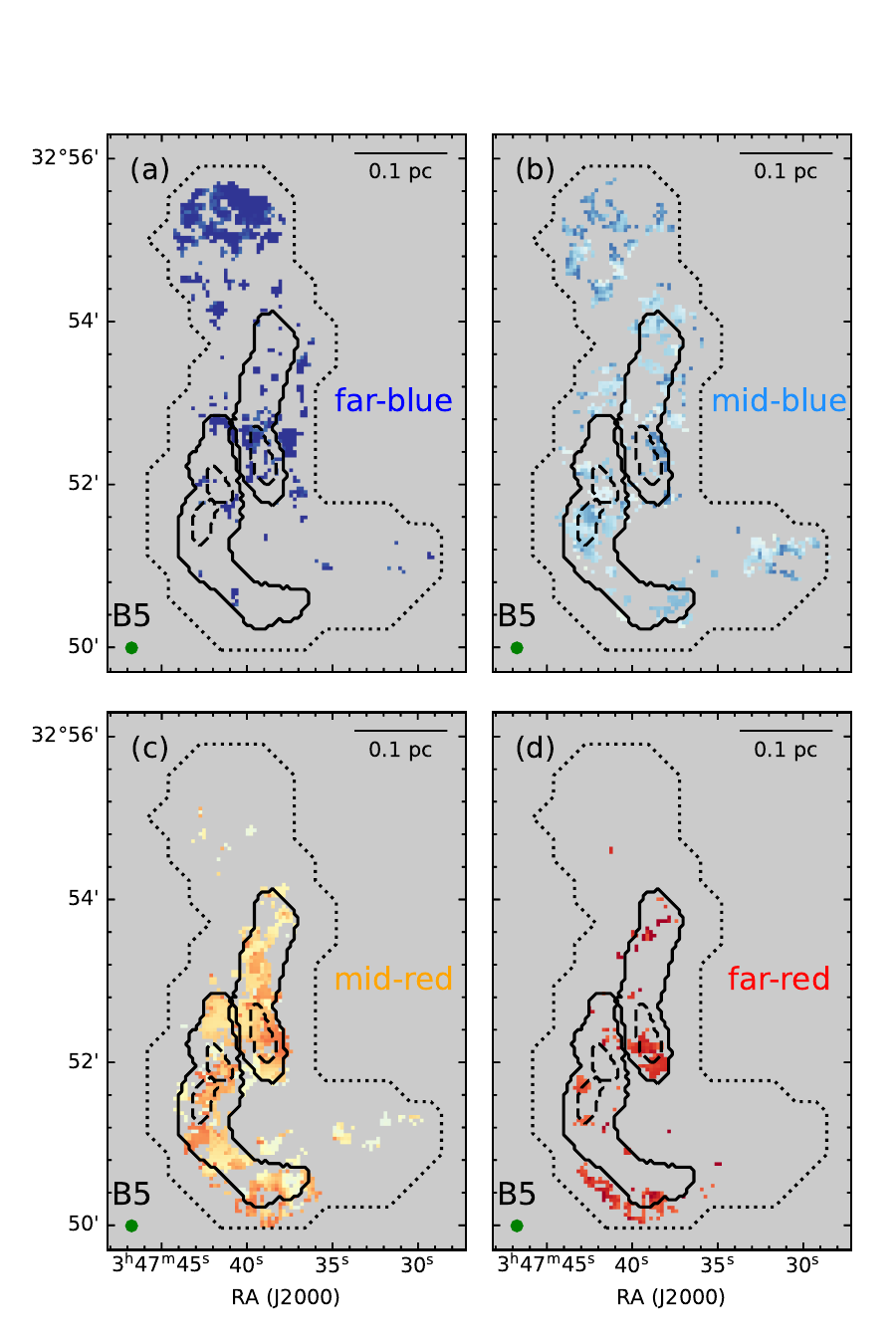}
  \caption{Velocity map of `global component' (left). Velocity map of the four additional components (right). 
  Each of these four components show a very narrow range of velocities. The median velocities of these components are 9.61 \kms (far-blue), 9.99 \kms (mid-blue), 10.32 \kms (mid-red), and 10.58 \kms (far-red).
  The black-solid and black-dashed contours show the two filaments and the three condensations in B5, respectively. 
  The common colour scale is shown in the left panel.
  The black-dotted contours show the boundary of the coherent core in \citet{pineda2010}. The scale-bar and beam size are shown in the top-right and  bottom-left corners, respectively.}
     \label{fig_vel_all_comp} 
\end{figure*}

\begin{figure*}[!ht]  
\centering
\includegraphics[width=0.50\textwidth]{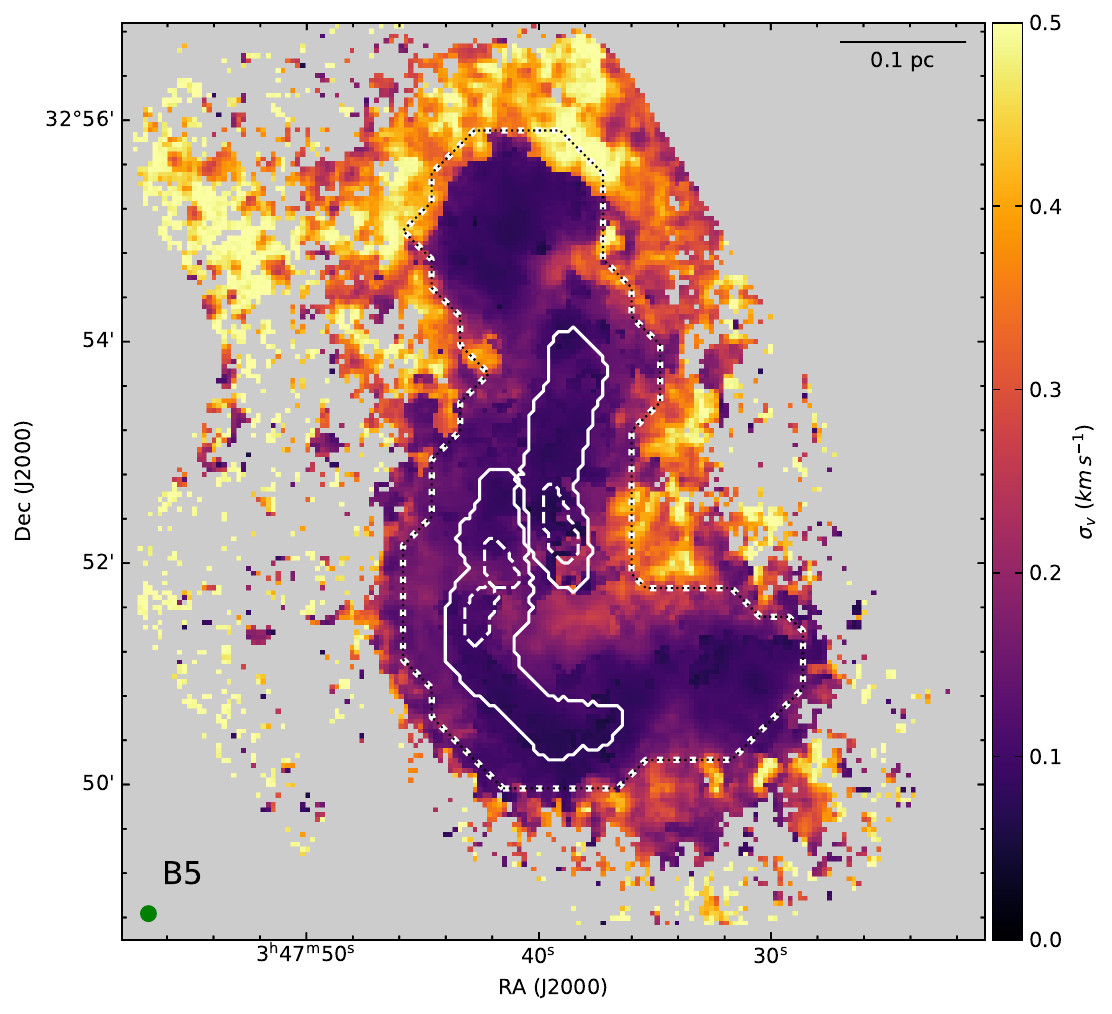}
\includegraphics[width=0.48\textwidth]{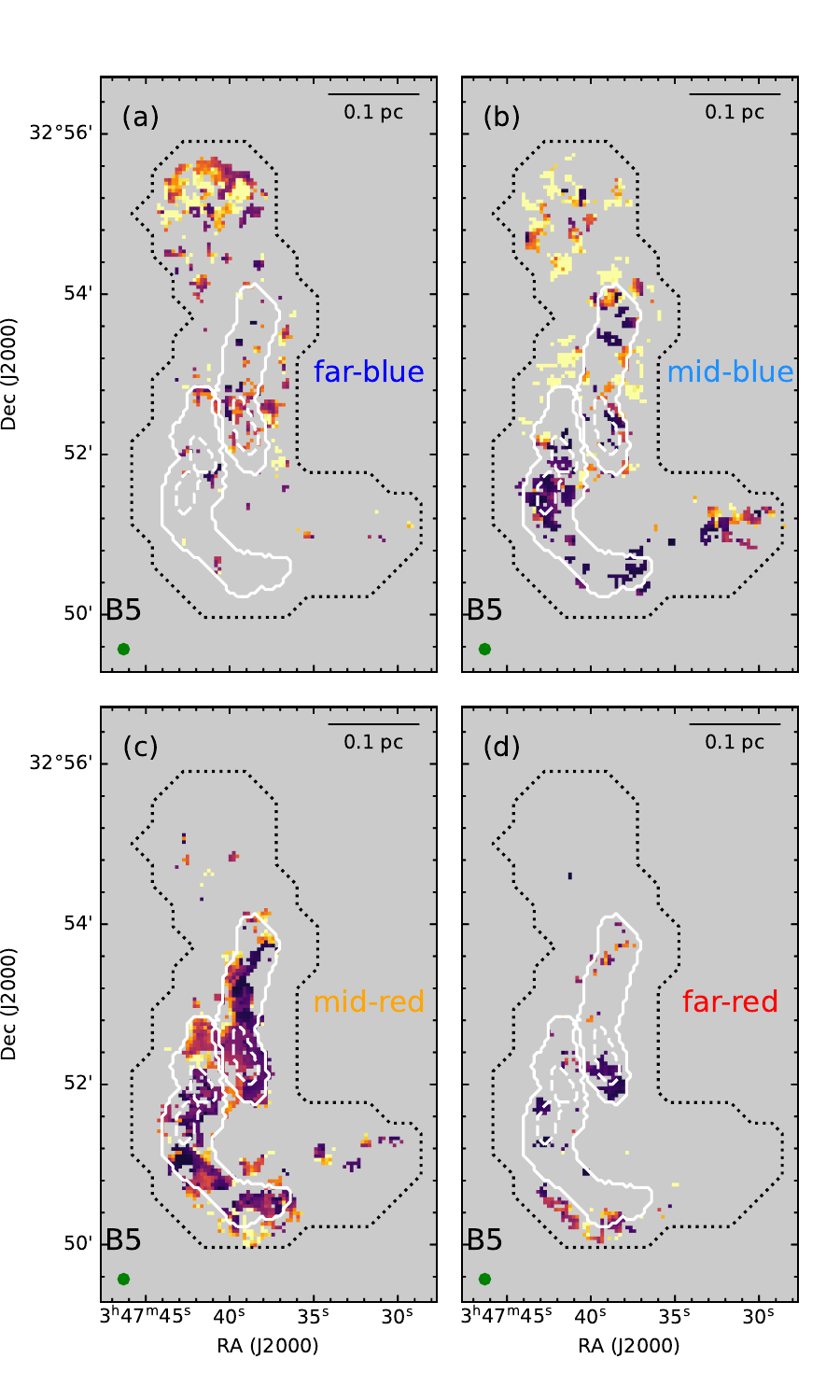}
  \caption{\textit{} Velocity dispersion of the `global component' (left). Velocity dispersions of the four additional components (right). The white-solid and white-dashed contours show the two filaments and the three condensations in B5, respectively. The white/black-dotted contours show the boundary of the coherent core in \citet{pineda2010}. The scale-bar is shown in the top-right corner and the beam size is shown in the bottom-left corner.}
     \label{fig_sig_all_comp} 
\end{figure*}

Figure \ref{fig_num_of_comp} shows the number of components fit towards different parts of B5.
Since each pixel is fitted independently, and the number of components varies across the map, the identification of spatially coherent features is non-trivial. 
With the exception of the two filaments and clump 1 (see Fig. \ref{fig_mom0}), we detected a single continuous component in the majority of the coherent core. We first isolated this component
with an approach similar to that of \citet{mike_2022_B5}. In regions with two or three components, we searched for the fitted component which provides the smoothest map of velocity and the brightness temperature, when combined with the single component in the outer region (see Appendix \ref{app_clust_meth} for details). This gives us a map of the component that is kinematically coherent over the largest spatial extent\footnote{The script used to sort the different components, as well as all other scripts and data files used in the analysis presented in this paper, can be accessed at \url{https://github.com/SpandanCh/Barnard5_filaments}}. 
Hereafter, this component is referred to as the `global component', as it is present throughout all the regions discussed in this work.
The velocity and velocity dispersion of this component are shown in the left panels in Figs. \ref{fig_vel_all_comp} and \ref{fig_sig_all_comp}. 
This component shows very similar properties as those presented in the single dish analysis of the region, which is more sensitive to the extended emission, \citep[see Fig. 2 in][]{pineda2010}.

We are then left with one or two more components in positions close to or towards the filaments. Upon visual inspection of these two components, we find that they are either red- or blue-shifted with respect to the median velocity of the global component ($\rm v_{ med} = 10.15\ \kms$). Therefore, we grouped the remaining two components into red-shifted and blue-shifted components. Each of these two components shows indications of further subdivision in velocity, and therefore, they are split into two further components each to give four kinematically coherent maps. The velocities and velocity dispersions of these four components are shown in the right panels of Figs. \ref{fig_vel_all_comp} and \ref{fig_sig_all_comp}. Due to insufficient sensitivities, we were not able to detect all of these components continuously, hence, these four maps appear patchy. 
We note that these different velocity components are not shown separately in Fig. \ref{fig_num_of_comp}, which only shows the number of components in the best-fit model at each pixel.

\citet{mike_2022_B5} presented a kinematical analysis of B5 using the same data set at the original resolution of 6\arcsec. However, they take a different statistical approach for the multi-component fit to the data using the \textit{MUFASA} software \citep{chen_mufasa}. Moreover, \citet{mike_2022_B5} attempted to fit up to only two velocity components in the region and focus their analysis on the one component that is kinematically coherent over the core region (very similar to `global component' in this work), while the other component is excluded.
In contrast, we focus our discussion on the four additional velocity components that we extract with the Bayesian approach to multi-component fit. 
Therefore, a direct comparison between the two works is not straightforward, even though both present an analysis of the signatures that hint at infalling material in the B5 filaments \footnote{Note: the definition of "filament"\ in \citet{mike_2022_B5} is also slightly different to that adopted in this work due to the methods employed.}.

\section{Results and discussions}
\label{sec_res}

\subsection{Identified velocity components}
\label{sec_vel_comp}

As can be seen from the left panel of Fig. \ref{fig_vel_all_comp}, the `global component' shows a smooth velocity field, fluctuating in the north-south direction. As this is also the brightest component, it dominates the velocity obtained from a single component fit to the data. Therefore, this velocity map is very similar to those obtained with a single component fit \citep{pineda2011_B5_fila,anika_2021}. 
Inside the core boundary \citep[from][dotted line in Fig. \ref{fig_vel_all_comp}]{pineda2010}, we attribute this component to the dense core material, and outside the core boundary, to the ambient cloud. 
This can be clearly seen from the velocity dispersion in the component (Fig. \ref{fig_sig_all_comp}). Inside the core boundary, the dispersion is subsonic ($\lesssim 0.2 \kms$) and outside the core, it is supersonic.

The four panels on the right of Fig. \ref{fig_vel_all_comp} show the velocities of the additional components we identify in B5. We refer to these components as `far-red', `mid-red', `mid-blue', and `far-blue' according to their shift in
velocities (as shown in the figure).
We are not able to detect all four components towards all positions, with a maximum of two (out of these four) components being detected at a given location. 
This is most likely due to insufficient sensitivity in the data. 

\begin{figure}[!ht]  
\centering
\includegraphics[width=0.4\textwidth]{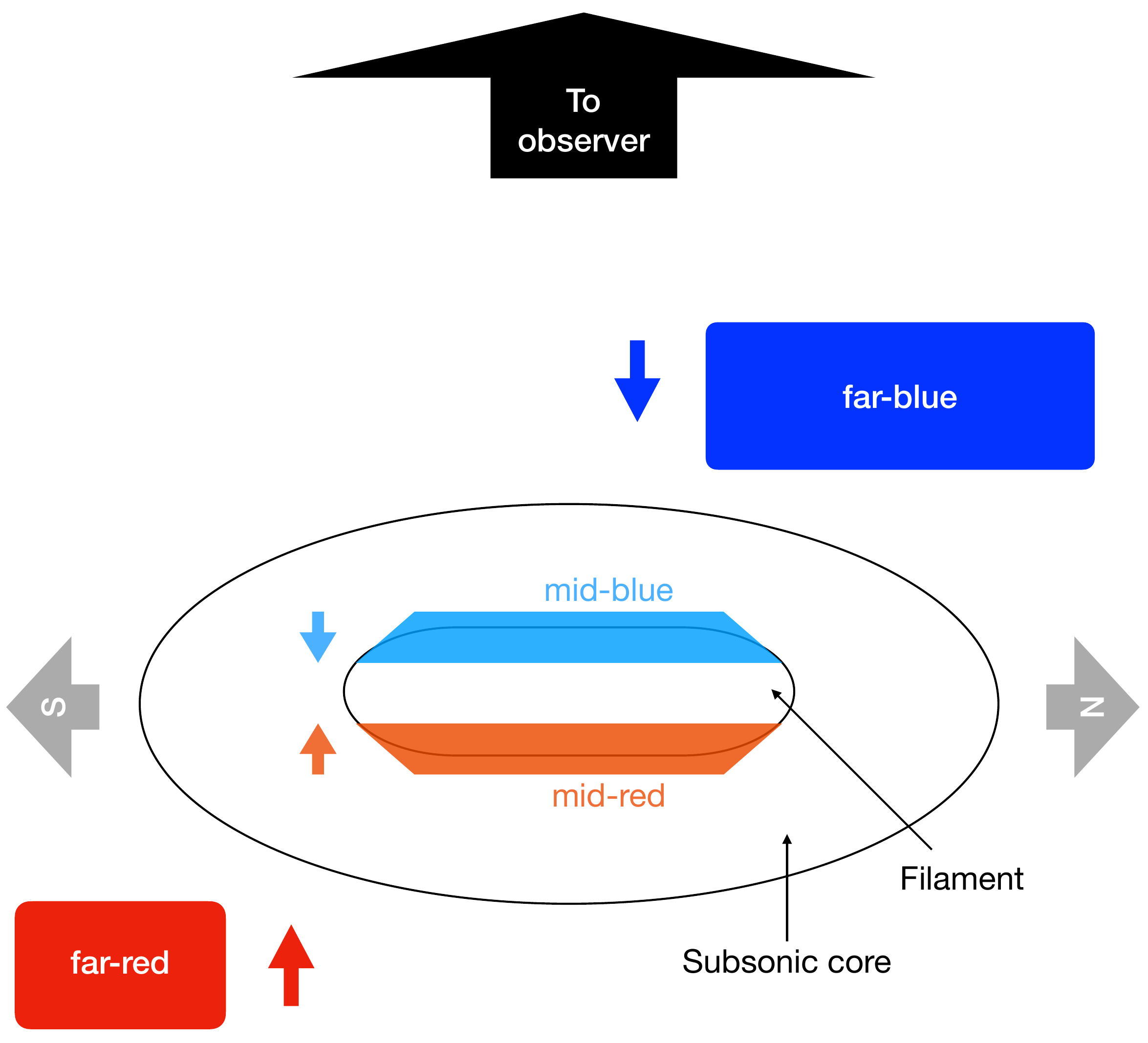}
  \caption{Schematic showing the spatial distribution of the different components in the line of sight. The coloured arrows show the direction of the velocity of the corresponding component. The ellipses represent the core and a filament inside the core, as indicated. The grey arrows indicate the north and south directions in the maps presented in this work.}
     \label{fig_cartoon_diff_comp} 
\end{figure}

The mid-blue and the mid-red components seem to trace roughly the two filaments. Each component shows almost constant velocities (median velocities of 9.99 \kms and 10.32 \kms, respectively) with a standard deviation less than 0.1 \kms. This suggests that each component is tracing a single, velocity coherent structure.
The velocity dispersion of these two components is subsonic inside the two filaments (Fig. \ref{fig_sig_all_comp}), which further shows that they are tracing material inside a subsonic core. Given the spatial extent and constant separation in velocity of these two components, their origin is most likely the two opposite sides of material undergoing infall onto the filaments (See Fig. \ref{fig_cartoon_diff_comp}).

The far-blue component shows an almost constant velocity around 9.6 \kms (standard deviation of 0.16 \kms), which indicates that the component is tracing a velocity coherent structure. The velocity dispersion in this component is largely supersonic (Fig. \ref{fig_sig_all_comp}), which suggests that this component traces gas in the ambient cloud, rather than the core. The velocity and the velocity dispersion of this component are similar to those of the global component in the north. Therefore, this component likely shows  material flowing from the ambient molecular cloud towards B5 from the north and north-west. We explore this possible mass infall in Section \ref{sec_infall_core}.
The location and velocity of this component is consistent with those of the methanol peak \citep[][observations with \emph{Herschel}, Onsala, and IRAM 30m observatories]{wirstrom2014_methanol_hotspot,taquet2017_methanol_B5}, which is indicative of gas collision. 
We note that there is a significant apparent gap in the map of this component between clump 1 and the central region. This could be due to some of the material in this component merging with the core material, making it difficult to distinguish from the core (global component). The lower velocity dispersions in the far-blue component towards the middle of the core, as compared to the northern part, could point to the turbulence in the component being dissipated in the process of merging with the core material.

The far-red component, which is red-shifted and at a velocity of approximately 10.6 \kms, could be largely divided into two parts: the arc-like structure just south of the boundary of filament-2, and region near condensation-1. The arc-like structure might show material falling into B5 from that direction similar to the aforementioned scenario towards north, as the velocities in this component are systematically different from that in the core and filament components (Fig. \ref{fig_vel_all_comp}). The region just south of condensation-1 might trace material falling towards the condensation due to gravitational collapse \citep[see also][]{mike_2022_B5}. A corresponding region is observed in the far-blue component north of condensation 1, which could be tracing the opposite side of this infall.
Figure \ref{fig_cartoon_diff_comp} shows the arrangement of the far-blue, mid-blue, mid-red and far-red components in the line of sight, as discussed in this section.

\subsection{Filament formation scenarios}
\label{sec_fila_form}

\citet{chen_2020_fila_form} proposed an analysis to determine whether filament formation is dominated by self-gravitational collapse or via direct shock compression. They define the dimensionless parameter \cv to differentiate between the two cases as 
\begin{equation}
    \label{eq_cv-chen}
    C_{\rm v} \equiv \frac{\Delta {\rm v}_h ^2}{G M(r)/L} ~,
\end{equation}
where $\Delta {\rm v}_h$ is half of the velocity gradient across the filament out to a distance, $r,$ from the centre, $G$ is the gravitational constant, and $M(r)/L$ is the mass per unit length of the filament at a location, $r$. Here, 
$\Delta {\rm v}_h^2$ is representative of the kinetic energy of
the flow transverse to the filament, while the denominator in Eq. \ref{eq_cv-chen} represents the gravitational potential energy in the filament. Therefore, the case where $C_{\rm v} \leq 1$ and $C_{\rm v} \gg 1$ represents scenarios where gravity or turbulence and flow dynamics, respectively, dominates the filament formation.

\begin{figure}[!ht]  
\centering
\includegraphics[width=0.3\textwidth]{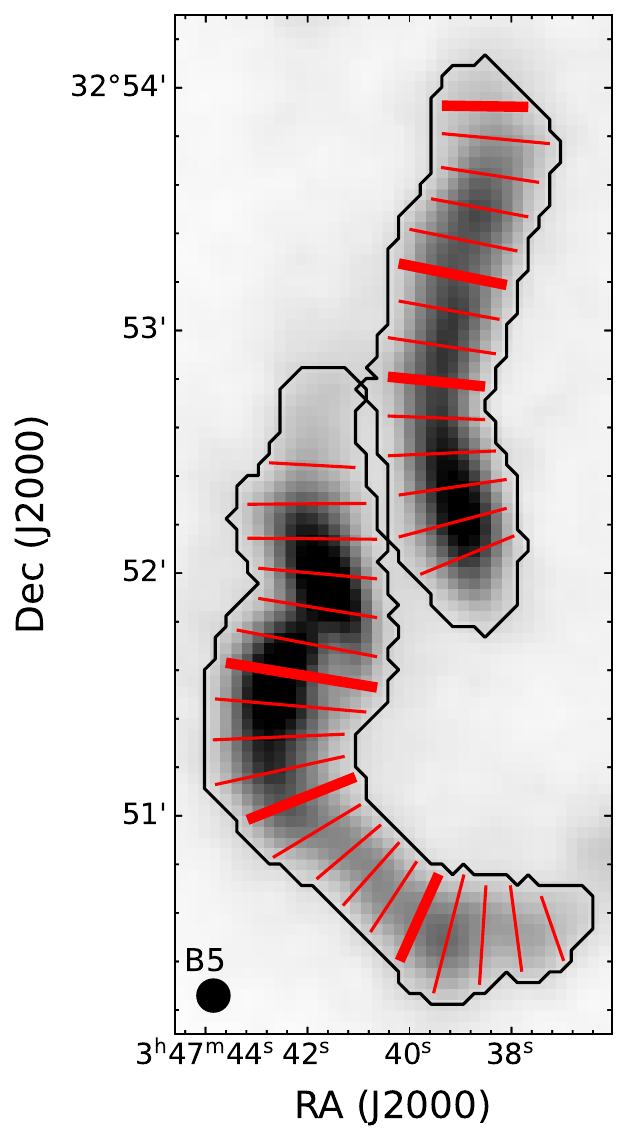}
  \caption{Orthogonal cuts to the major axes of the two filaments. The cuts corresponding to the velocity gradients shown in Fig. \ref{fig_vel_grad_cv_fil_exmpl} are shown with thick lines. The background greyscale shows the integrated intensity of \amm (1,1). The cuts are spaced one beam apart. The beam is shown in the bottom-left corner.}
     \label{fig_spine_cuts} 
     
\end{figure}

We determine the boundaries and the spine of the filaments following the analysis of \citet{anika_2021}. We define orthogonal cuts to each of the filament spines with a separation of one beam between each cut. These cuts are shown in Fig. \ref{fig_spine_cuts}. For each of these cuts we calculated the difference in the velocity at the two ends (at the filament boundaries) and the mass per unit length. 
As described in Section \ref{sec_vel_comp}, the mid-red and mid-blue components likely trace infall onto the filaments in the line of sight. For this analysis, we needed to measure a velocity gradient in the plane of sky, both inside and in the neighbourhood of the filaments.
Therefore, we used the map of the global component (left panel in Fig. \ref{fig_vel_all_comp}) here.
We calculated the mass using an \amm\ intensity to mass conversion factor of 1.5$\pm$0.7\,\msun\ (Jy$\,$beam$^{-1}$)$^{-1}$, derived in \citet{anika_2021} by comparing the \amm\ (1,1) emission and the 450 $\rm \mu$m continuum flux from SCUBA-2 \footnote{Submillimetre Common-Use Bolometer Array 2 at the James Clerk Maxwell Telescope (JCMT).}.

\begin{figure*}[!ht] 
\centering
\includegraphics[width=0.9\textwidth]{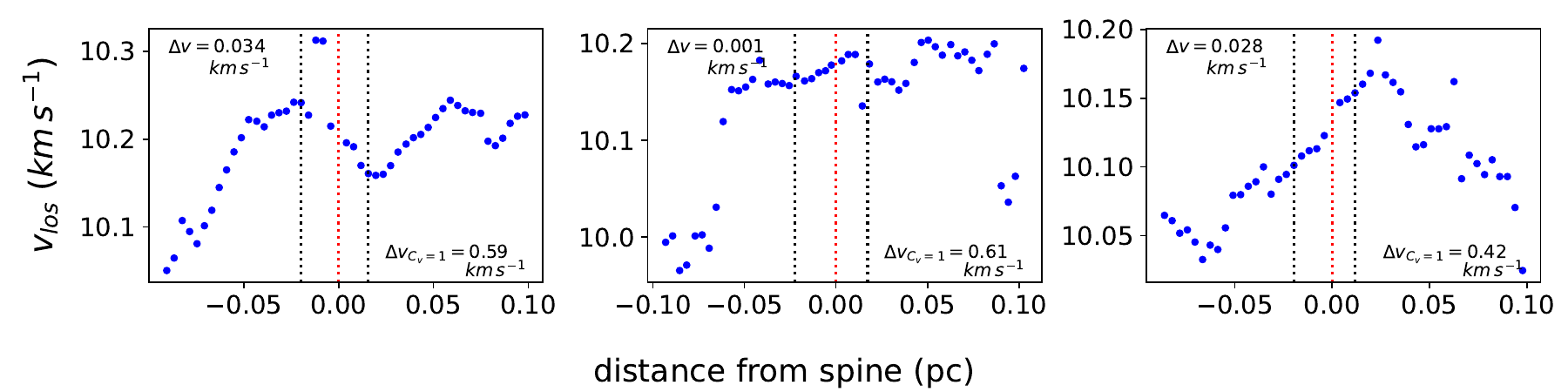}
\includegraphics[width=0.9\textwidth]{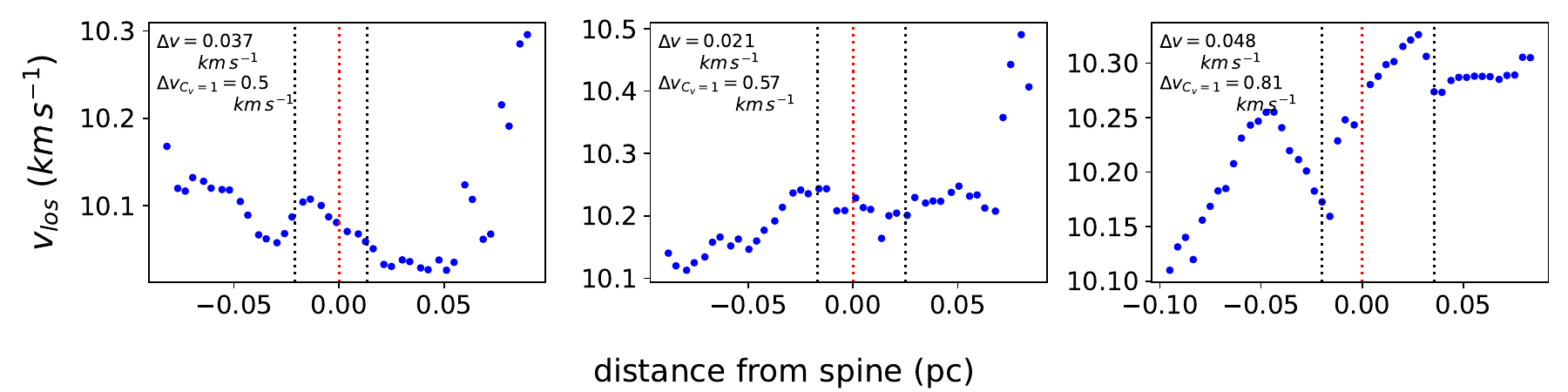}
  \caption{Selected examples of the line of sight velocity profiles in the orthogonal cuts in filament-1 (top) and filament-2 (bottom). The red-dotted vertical lines show the position of the filament spine and the black-dotted vertical lines show the boundary of the filament for the respective orthogonal cuts shown. 
  The $\Delta {\rm v}$ parameter ($\Delta {\rm v}_h$ in Eq. \ref{eq_cv-chen}) for the cut is shown in the upper-left corner. The $\Delta {\rm v}_h$ value needed to have $C_{\rm v} = 1$ ($\Delta {\rm v_{C_V=1}}$) is shown in the bottom-right corner for the upper panels, and below the $\Delta {\rm v}_h$ in the lower panels.
  }
     \label{fig_vel_grad_cv_fil_exmpl} 
     
\end{figure*}

Figure \ref{fig_vel_grad_cv_fil_exmpl} shows six selected examples of the gradients in velocity across the orthogonal cuts to the filaments 1 and 2. The corresponding cuts are shown with thick lines in Fig. \ref{fig_spine_cuts}. The gradients in velocity across all the cuts are shown in Appendix \ref{app_vel_ortho_cut_all}. 
Overall, we find \cv values less than 0.03, and so, across these two filaments, $C_{\rm v} \ll 1$ (Fig. \ref{fig_kde_cv}).

The average $\Delta {\rm v}_h$ for the orthogonal cuts along the two filaments is 0.04 \kms. For the \cv values to be near unity, $\Delta {\rm v}_h$ needs to be $\approx$0.5 \kms, on average, across the cuts. Therefore, the observed $\Delta {\rm v}_h$ values are more than ten times lower than what is needed for $C_{\rm v} = 1$ at the mass-per-unit-length across the filaments. For comparison, the sound speed at the temperatures around the filaments is smaller than 0.2 \kms. The gas in that region is subsonic, therefore, the observed velocity dispersions are even smaller. By using the components that are likely tracing infall onto the filaments in the line of sight (mid-red and mid-blue), we get $\Delta {\rm v}_h \approx 0.15\, \kms$, which is also much lower than what is required to have $C_{\rm v} = 1$.

As mentioned by \citet{chen_2020_fila_form}, filaments formed in a planar-geometry (preferred direction of accretion) could have $C_{\rm v} \ll 1$ due to projection effects or if the filaments are at an early formation stage. The latter scenario is consistent with the prestellar condensations, but is not consistent with the presence of the Class 1 protostar in B5, nor the highly supercritical $M/L$ of the filaments.
In the former scenario, if the slab containing the filament is inclined at an angle, $\theta,$ with the plane of the sky, the observed velocity difference across the filament, $\Delta {\rm v}_{h,obs}$ will be
\begin{equation}
    \Delta {\rm v}_{h, obs} = \Delta {\rm v}_h\ sin \theta ~,
\end{equation}
where $\Delta {\rm v}_h$ is the true velocity difference.
The true \cv value would then be $C_{{\rm v}, obs}/sin^2 \theta$, where $C_{{\rm v}, obs}$ is what we report here. Considering the \cv values we obtain across the filaments, the inclination angle needs to be $|\theta| \lesssim 7 ^\circ$, for them to be comparable to the values obtained by \citet{chen_2020_fila_form}, considering that $|\theta| = 45 ^\circ$ in their results. Therefore, it is very unlikely for the estimated low values of \cv to be entirely due to inclination effects.

Since we observe $C_{\rm v} \ll 1$ across the two filaments, we can rule out the turbulence/flow formation scenario. \citet{anika_2021} find mass-per-unit-length ($M/L$) in these two filaments to be much higher ($\sim$ 80\,\msun\ pc$^{-1}$) than what can be thermally supported ($\sim$ 17\,\msun\ pc$^{-1}$).
Therefore, both the filaments are strongly dominated by gravity. However, we do not see a corresponding strong gradient in the velocity field, which suggests the presence of additional support against gravity.
A strong poloidal magnetic field that is parallel to the spine of the filament could provide this necessary support. 
\citet{anika_2021} estimated the required magnetic field strength to be $\sim 500\, \rm\mu$G. The estimated Alfven velocity, $V_A$\ (=B/$\sqrt{4\pi\rho}$, where B is the magnetic field strength, and $\rho$ is the gas mass density) in the filaments is $\approx 0.5\, \kms$, which is almost a factor of three higher than the observed velocity dispersions in the filaments (see Fig. \ref{fig_sig_all_comp}). 
This estimated value of $V_A$ is similar, however, to the $\Delta {\rm v}_h$ required to have $C_{\rm v} = 1$; that is, the scenario where gravitational potential energy is comparable to the kinetic energy in the filaments. Therefore, it supports the additional support against gravity being provided by the estimated magnetic field.

We note that the values for \cv in our work are much smaller than those for the models considered by \citet{chen_2020_fila_form}, who reported $C_{\rm v} \sim 0.3$. This suggests that their simulations are not good representations of the filaments in B5. We discuss three possible factors behind the significant difference in the \cv values between our work and the simulations of \citet{chen_2020_fila_form} in Appendix \ref{app_chen_compare}.

\subsection{Infall in the filaments}
\label{sec_infall_fila}

Assuming that both mid-blue and mid-red velocity components (see Sect. \ref{sec_vel_comp}) trace the two opposite sides of infalling material, we can estimate the rate of infall onto the two filaments. The infall rate is then given by: 
\begin{equation}
\label{eq_infall}
    \dot{m} = S \times \rho \times {\rm v}_{infall} ,
\end{equation}
where \textit{S} is the surface area of the filament, $\rho$ is the volume density inside the core, and ${\rm v}_{infall}$ is the velocity of the infall, taken to be the mean difference between the red- and blue-shifted filament components (mid-red and mid-blue). 
As both the red- and blue-shifted filament components (mid-red and mid-blue components, respectively) have roughly constant velocities of 10.3 \kms\ and 9.9 \kms, respectively, we take 
\begin{equation}
    {\rm v}_{infall} = \frac{<{\rm v}_{red} - {\rm v}_{blue}>}{2} = 0.2\ \kms ~.
\end{equation}

\begin{figure}[!ht]  
\centering
\includegraphics[width=0.5\textwidth]{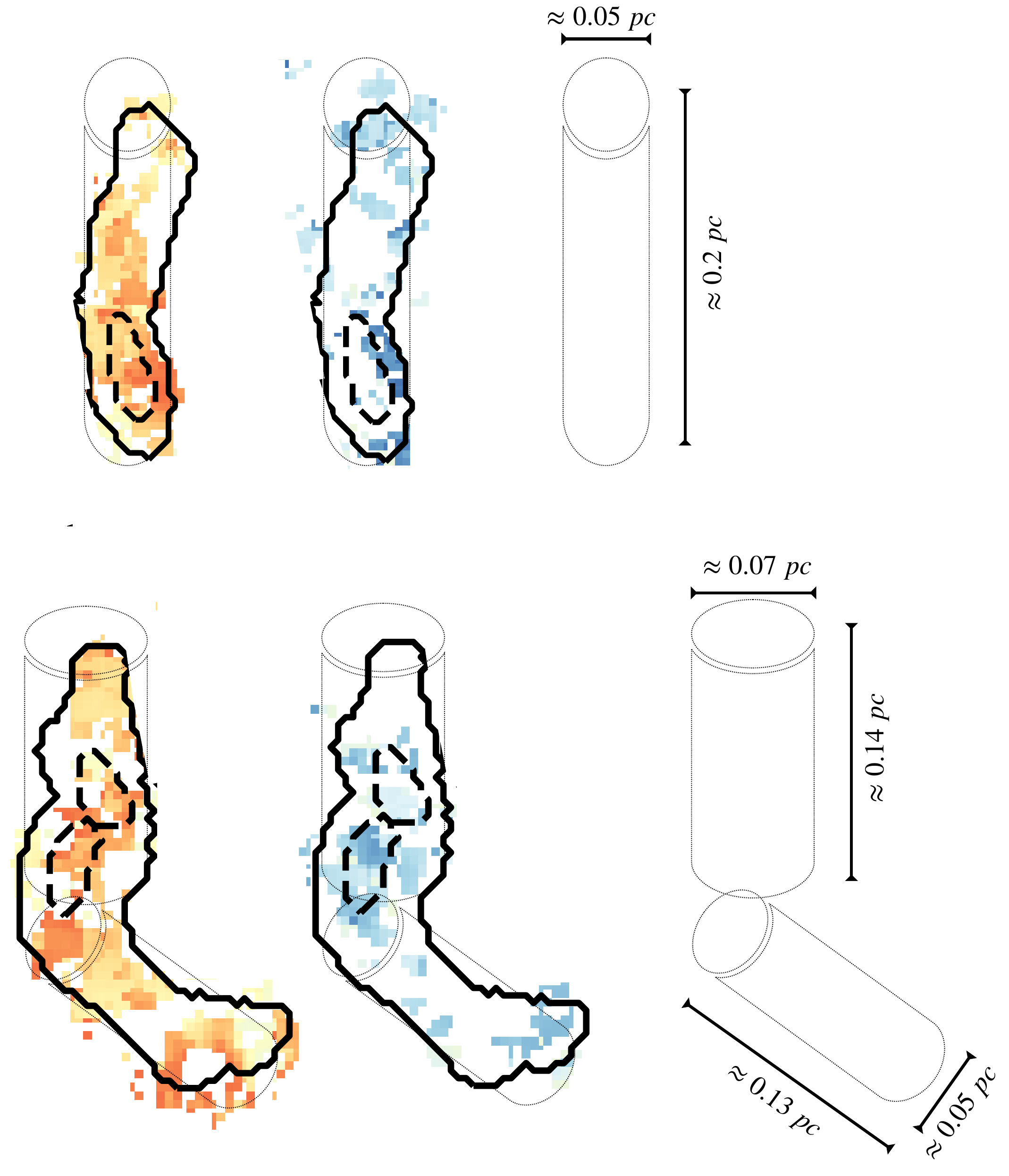}
  \caption{Geometry considered to calculate the mass infall onto filament-1 (top) and filament-2 (bottom).}
     \label{fig_infall_fila1} 
     
\end{figure}
\FloatBarrier

Figure \ref{fig_infall_fila1} shows a schematic of the geometry considered to calculate the approximate surface area of the two filaments. In the case of filament-1, we estimate the surface area assuming the filament to be roughly cylindrical with a diameter $d$=0.05 pc and a length $l$=0.2 pc. The surface area then is $S=2\pi(\frac{d}{2})l$. For filament-2, we estimate the surface area by dividing the filament into two roughly cylindrical regions with diameters $d_1$=0.07 pc and $d_2$=0.05 pc, and with lengths $l_1$=0.14 pc and $l_2$=0.13 pc, respectively. The total surface area is $S = S1+S2 = 2\pi(\frac{d_1}{2})l_1 + 2\pi(\frac{d_2}{2})l_2$.

Using the filament masses from \citet{anika_2021} and the geometry described above, we obtain a volume density
of $\rm n_{H_2} = 3.5\times10^5\,$\cc\ for filament-1 and $\rm n_{H_2} = 2.5\times10^5\,$\cc\ for filament-2. As the different components are in the same line of sight, the exact densities associated with the components are difficult to distinguish. Therefore, we use the average density of the filament to calculate the infall rate.
With these estimates, we get infall rates of $1.6\times10^{-4}\, \msunyr$ for filament-1 and $1.8\times10^{-4}\, \msunyr$ for filament-2.
The total masses of the two filaments are 9.4$\,$\msun\ and 13.4$\,$\msun, respectively \citep{anika_2021}.
At these rates, the formation timescale of the two filaments, as well as the mass doubling time, is 6-7$\times 10^4$ yr, which is an order of magnitude less than typical core lifetimes of 0.5-1 Myr \citep{stella_2022_core_evo}. This implies that at these infall rates, the filament masses can change significantly during the evolution of the core.
As the material likely undergoing infall is at the surface, the density is expected to be lower than the average density, and hence, these estimates are upper-limits for the infall rates. 
Using observations with HC$_3$N (which traces chemically fresh gas, as compared to \amm; \citealt{suzuki_1992}) and comparison with the \amm data, \citet{teresa_b5_infall} suggest infall of fresh material from the B5 core onto the filaments. This could correlate to the infall traced by the mid-red and mid-blue components, as discussed above.

\subsection{Mass infall onto B5}
\label{sec_infall_core}

\begin{figure}[!ht]  
\centering
\includegraphics[width=0.3\textwidth]{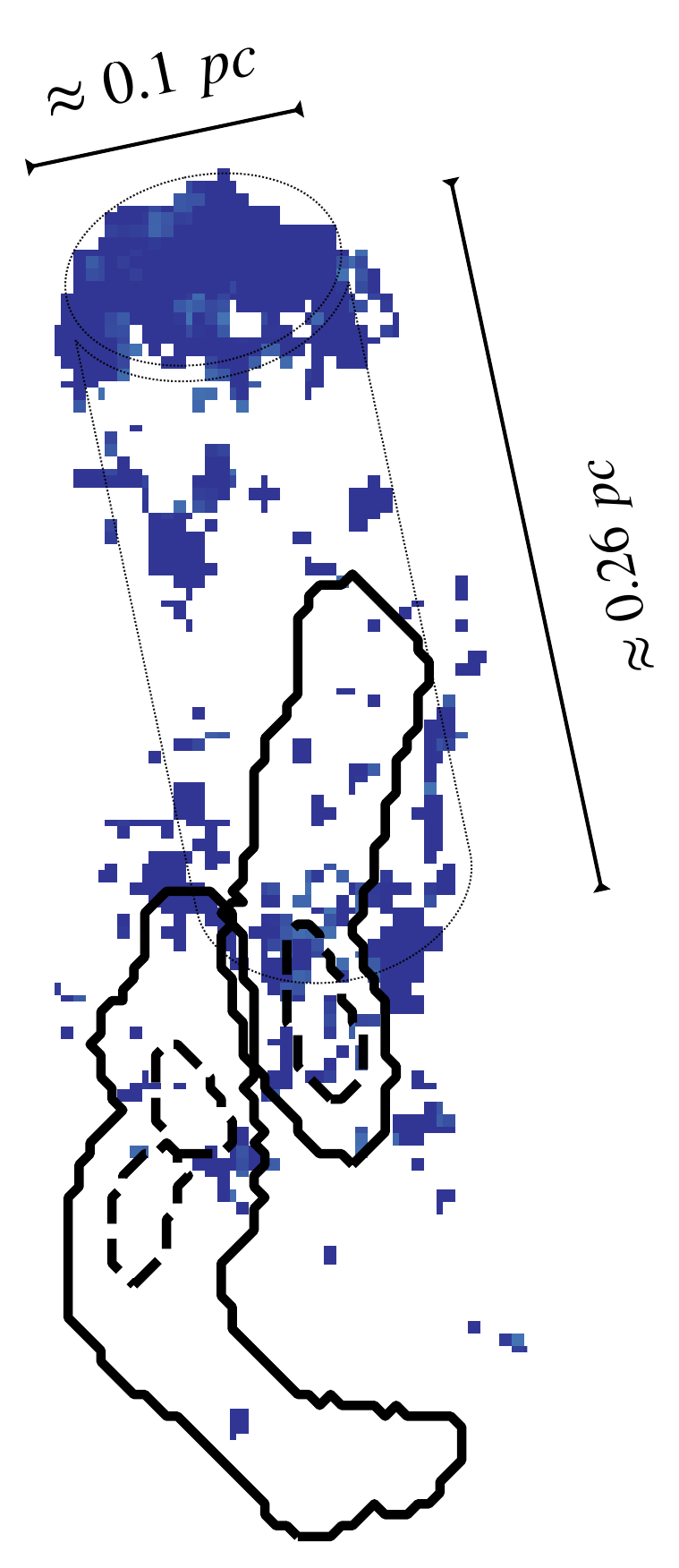}
  \caption{Geometry considered for calculating the mass infall onto B5.}
     \label{fig_infall_B5} 
\end{figure}

Assuming that the far-blue component (top left panel on the right of Fig. \ref{fig_vel_all_comp}) traces material falling onto B5, we can similarly estimate its rate of infall. We assume that this material is flowing through a cylindrical region shown in Fig. \ref{fig_infall_B5}, although only one-third of it (approximately) is detected\footnote{For this component, the relatively poor coverage is presumably due to lack of sensitivity (see Sect. \ref{sec_vel_comp}).}.

\FloatBarrier

We estimated the rate of infall of material onto B5, using Eq. \ref{eq_infall}. Again, as this component is in the same line of sight as the other components, it is difficult to obtain the exact density associated to the material. So, we assumed a volume density 
of $\rm n_{H_2}\sim 10^4\, \cc$, characteristic for core material, for the infall rate calculation.
As the mean velocity in the core around this region is $\approx 10.2\, \kms$ (Fig. \ref{fig_vel_all_comp}), we take the infall velocity to be 0.6 \kms\ (as the far-blue material is at ${\rm v}=9.6\, \kms$). We then obtain a mass infall rate of ${3.5\times10^{-5}\, \msunyr}$ onto B5. We note that as this component is sparsely sampled, this is an upper limit of the mass infall rate. 

Using observations with HC$_7$N, \citet{friesen_2013_fila_infall} calculated a lower limit of $2-5\times10^{-6}\, \msunyr$ for accretion onto dense filaments in the Serpens South, which is a cluster-forming region. As HC$_7$N traces a lower density compared to \amm, the infall observed by \citet{friesen_2013_fila_infall} is likely 
in a different, more external layer in the cloud, than what would be traced by \amm.
The observation of different regions of mass infall with comparable rates might point to the existence of a mechanism through which new material is being delivered to coherent cores. In \citet{paper_I}, we also discuss possible material accretion onto coherent cores. These results highlight the possibility of observing infall onto cores in different star-forming regions, using suitable tracers of different densities. If accretion of new material onto the core is observed towards more regions, there could be a need for a significant update to the current core evolution models, where cores are assumed to evolve in isolation.

\section{Conclusions}
\label{sec_conclu}

We used combined GBT and VLA observations of the star-forming core Barnard 5 in Perseus to study the multi-component structure of the core. Our results can be summarised as follows:

\begin{enumerate}
    \item By smoothing the data to a beam of 8$''$ (the original beam was 6$''$), we are able to identify different velocity components in the line of sight. Apart from the brightest component (which has properties similar to those from a single component fit), we detected four separate velocity components. We label these four components `far-red', `mid-red', `mid-blue', and `far-blue', alluding to their redshift or blueshift relative to the average velocity of the cloud.
    
    \item The mid-red and mid-blue components are separated by 0.4 \kms and they are detected primarily towards both filaments in B5.
    We postulate that these two components likely trace the two opposite sides in the line of sight of material contracting within the filaments.
    
    \item We find that the simulations presented in \citet{chen_2020_fila_form} exploring filament formation scenarios do not accurately represent the filaments in B5. For both filaments in B5, we find $C_{\rm v} \ll 1$ across all the orthogonal cuts, and therefore, we rule out the turbulence- and flow-dominated formation scenarios for the two filaments. Both filaments have highly supercritical masses per unit of length, but do not show any strong orthogonal velocity gradient. This indicates a presence of additional support against gravity, likely via a poloidal magnetic field that is parallel to the spine of the filaments.
    
    \item Assuming simple geometry and uniform density inside the core, we estimated the rate of infall of material onto the two filaments (seen with mid-red and mid-blue components) to be $\rm 1.6\times10^{-4}\, \msunyr$ and $\rm 1.8\times10^{-4}\, \msunyr$, respectively, for filament-1 and filament-2. Since the density at the surface, where the accretion likely takes place, is lower than the average density in the core, we note that these rates are upper-limits for the infall. The mass doubling time at these infall rates is 6-7$\times 10^4$ yr. Therefore, the filament masses can change significantly during the typical core lifetime of 0.5-1 Myr.
    
    \item We also obtain an upper limit of $\rm 3.5\times10^{-5}\, \msunyr$ for the rate of infall of mass onto B5 (using the far-blue component) from the north. This rate is comparable to the rate of infall onto dense filaments in Serpens South. If more cores are observed to accrete new material after their formation, that would indicate that a significant update is required in the  current core evolution models, where the cores are treated as  evolving in isolation.

    \item Our results highlight that infall at different scales within dense cores can be traced with high-sensitivity observation of suitable molecular lines using multi-component analysis. Observations of accretion of fresh new material onto dense cores during its evolution suggest that the current physical and chemical models of core evolution might need a significant update.
\end{enumerate}

\begin{acknowledgements}
SC, JEP, PC and MTV acknowledge the support by the Max Planck Society.
\end{acknowledgements}

\bibliographystyle{aa}
\bibliography{biblio}

\begin{appendix}

\section{Maps of Bayes factor $K^a_b$}
\label{app_k_factor}

\begin{figure}[h!t]  
\centering
\onecolumn
\includegraphics[width=0.48\textwidth]{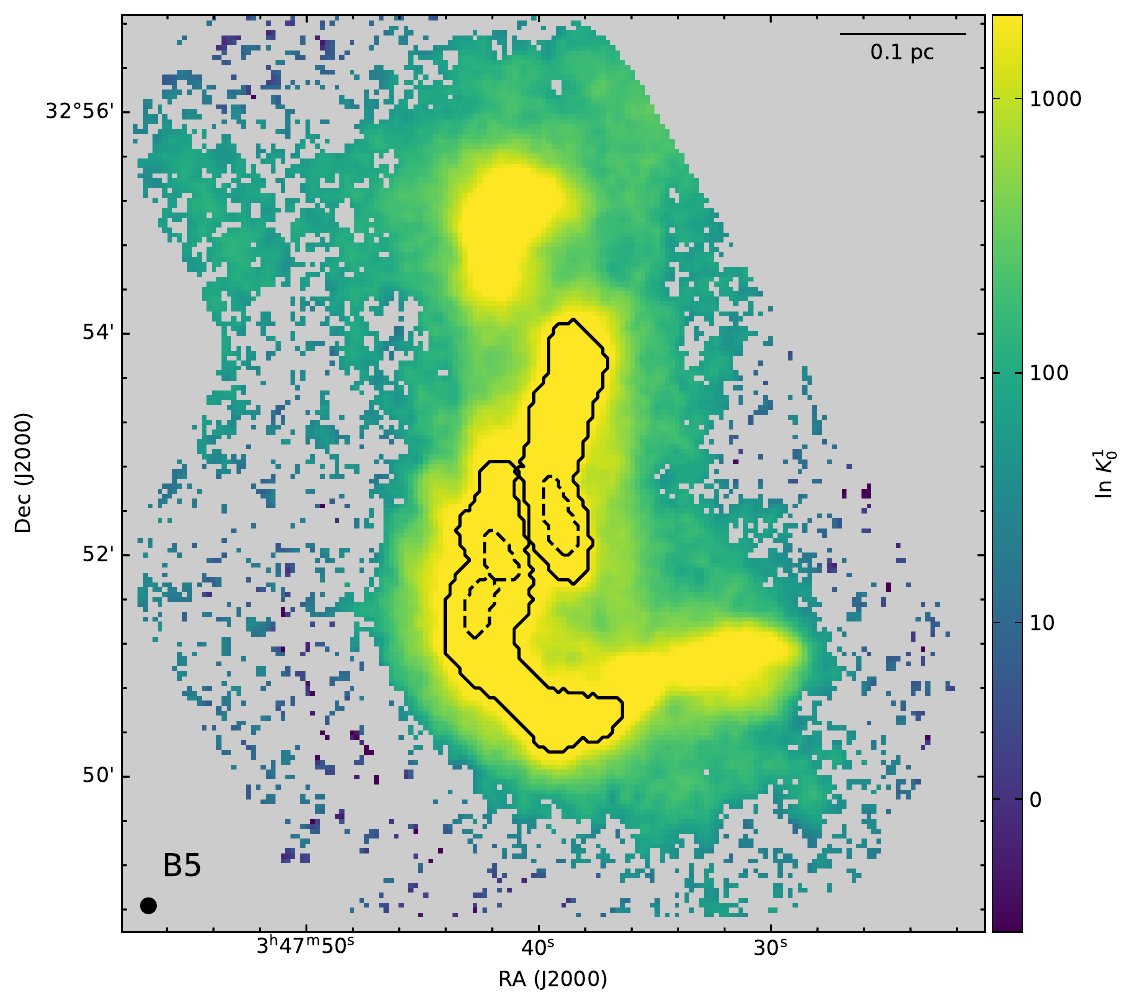}
\includegraphics[width=0.48\textwidth]{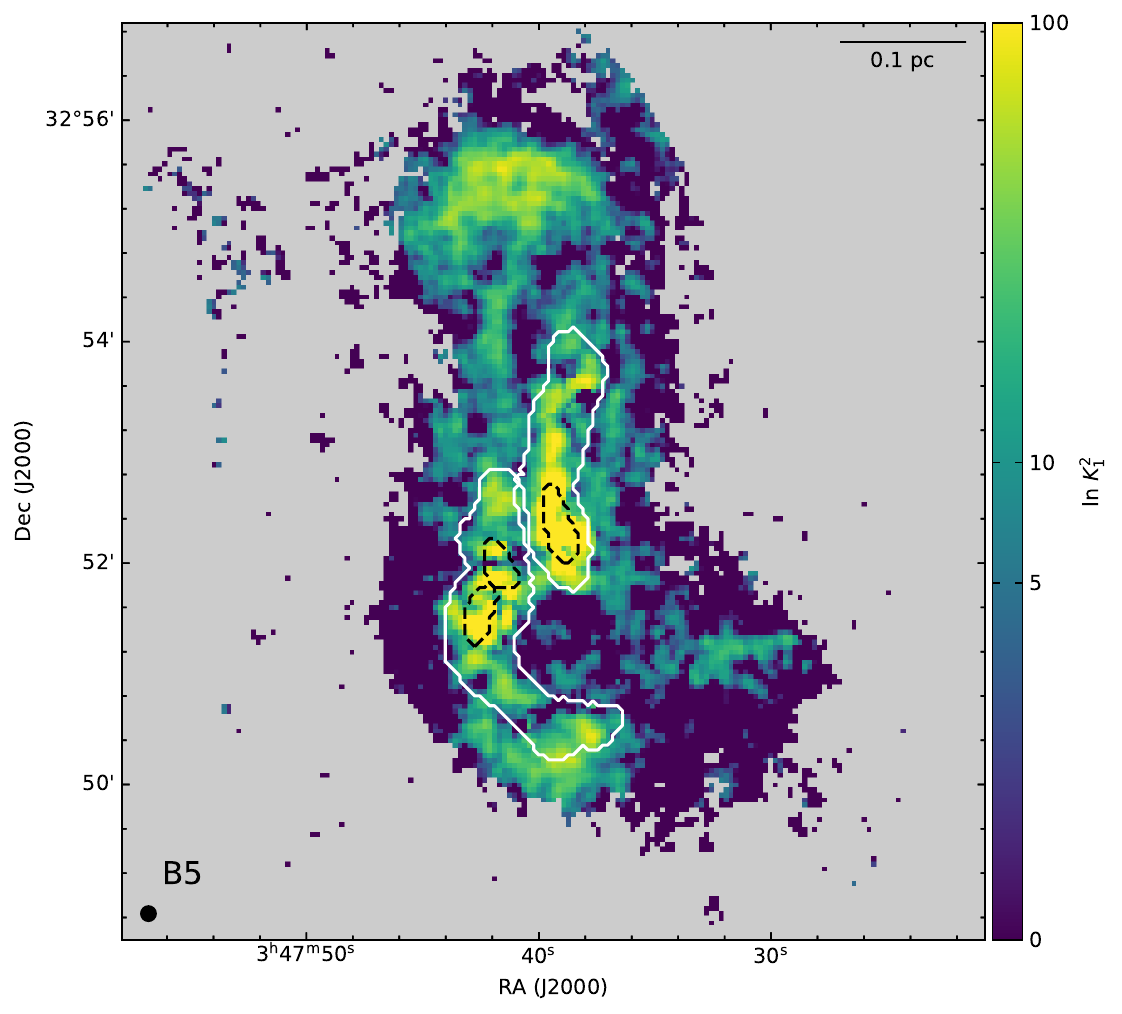}
\includegraphics[width=0.48\textwidth]{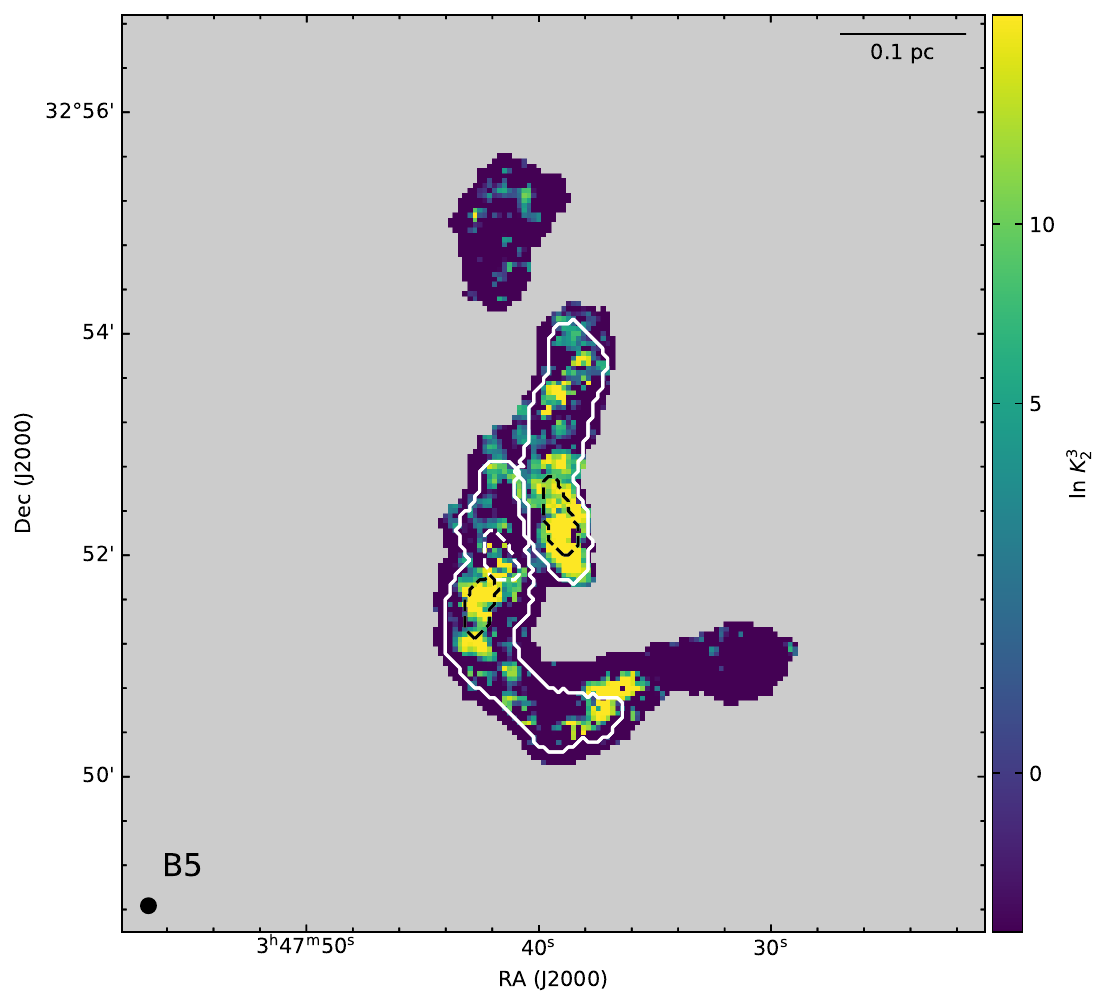}
\caption{Maps of the Bayes factors, $K^1_0$ (top left), $K^2_1$ (top right), and $K^3_2$ (bottom). The black-solid or white-solid contours show the two filaments and black-dashed or white-dashed contours show the three condensations in B5, respectively. The scale-bar and beam size are shown in the top-right and  bottom-left corners, respectively.}
     \label{fig_k-factors} 
\end{figure}

As mentioned in Sect. \ref{sec_line_fit}, we use the Bayes factor, $K^a_b$ (given by Eq. \ref{eq_k_factor}), to decide whether model $a$ is favoured over model $b$. Figure \ref{fig_k-factors} show the maps of the Bayes K-factors calculated between no fit and a 1-component fit ($K^1_0$), between 1- and 2-component fits ($K^2_1$) and between 2- and 3-components ($K^3_2$). Following the work of \citet{vlas_2020_bayesian}, we consider a threshold of $ln\, K^a_b \ge 5$ to select model $a$ over model $b$.

\FloatBarrier
\FloatBarrier

\begin{multicols}{2}

\section{Clustering method used in component assignment}
\label{app_clust_meth}

As mentioned in Sect. \ref{sec_comp_sort}, in the pixels with more than one velocity component in the best fit model, we select the component that results in smooth maps of centroid velocity ($\vel$) and the brightness temperature ($T_{MB}$). In other words, our grouping is done to minimise sudden and sharp changes in these two parameters between nearby pixels. For this purpose, we define a weighted dimensionless parameter distance, $d_p$, between two pixels as

\begin{equation}
    d_p = \sqrt{\left ( w_v \times \frac{\Delta \vel}{\rm v_{norm}} \right )^2+ \left( (1 - w_v) \times \frac{\Delta \rm T_{MB}}{\rm T_{MB, norm}} \right)^2} ~,
\end{equation}

where $\Delta \vel$ and $\Delta \rm{T_{MB}} $ are the differences in the two parameters at the two pixels, and $w_v$ is the fractional weight ($0 < w_v < 1$) applied to $\Delta \vel$. 
The separations in \vel and $\rm T_{MB}$ are normalised by $\rm v_{norm}$ and $\rm T_{MB, norm}$, respectively, which are the maximum differences in the respective parameters in the entire region. We use $\rm v_{norm} = 1\ \kms$ and $\rm T_{MB, norm} = 7.5\ K$ (calculated from one-component fit results).
Starting with a pixel in the region with a one-component fit, we look at each of the neighbouring pixels, search for the component with the lowest $d_p$, and group that component to the one-component maps. We then move radially outward, and repeat this process for the neighbours of each new pixel, until the entire region is covered. 
Thus, we obtain the extended component, which we refer to as `global component'.
We explore clustering with different relative weights in the parameters ($w_v$ = 0.2-0.8) and different starting points, but find that it does not significantly change the maps of global component.

We then have one or two additional components in some pixels. Using different clustering methods as mentioned above, we are unable to get smooth parameter maps for these two components, which indicates the need for further subgroups.
The additional components are then grouped according to their velocity, as described in Sect. \ref{sec_comp_sort}.

\section{Line of sight velocities along orthogonal cuts through major axes of the filaments}
\label{app_vel_ortho_cut_all}

\begin{figure}[h!t] 
\centering
\includegraphics[width=0.97\textwidth]{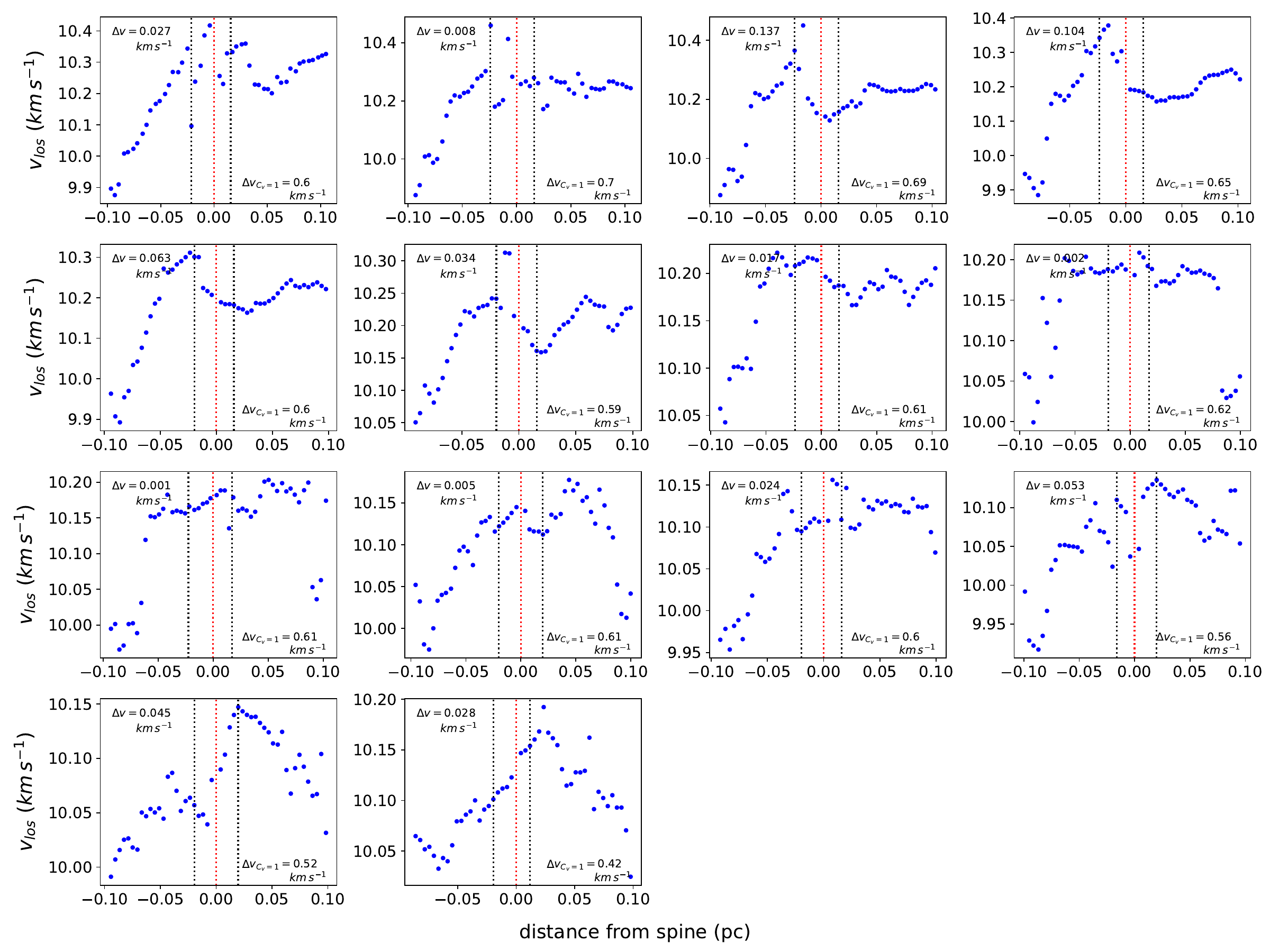}
  \caption{Line of sight velocity profiles in the orthogonal cuts in filament 1. The red-dotted vertical lines show the position of the filament spine and the black-dotted vertical lines show the boundary of the filament for the respective orthogonal cuts shown. 
  The $\Delta {\rm v}$ parameter ($\Delta {\rm v}_h$ in Eq. \ref{eq_cv-chen}) for the cut is shown in the upper-left corner. The $\Delta {\rm v}_h$ value needed to have $C_{\rm v} = 1$ ($\Delta {\rm v_{C_V=1}}$) is shown in the bottom-right corner.
  }
     \label{fig_vel_grad_cv_fil1} 
     
\end{figure}

\begin{figure}[!ht]  
\centering
\includegraphics[width=0.9\textwidth]{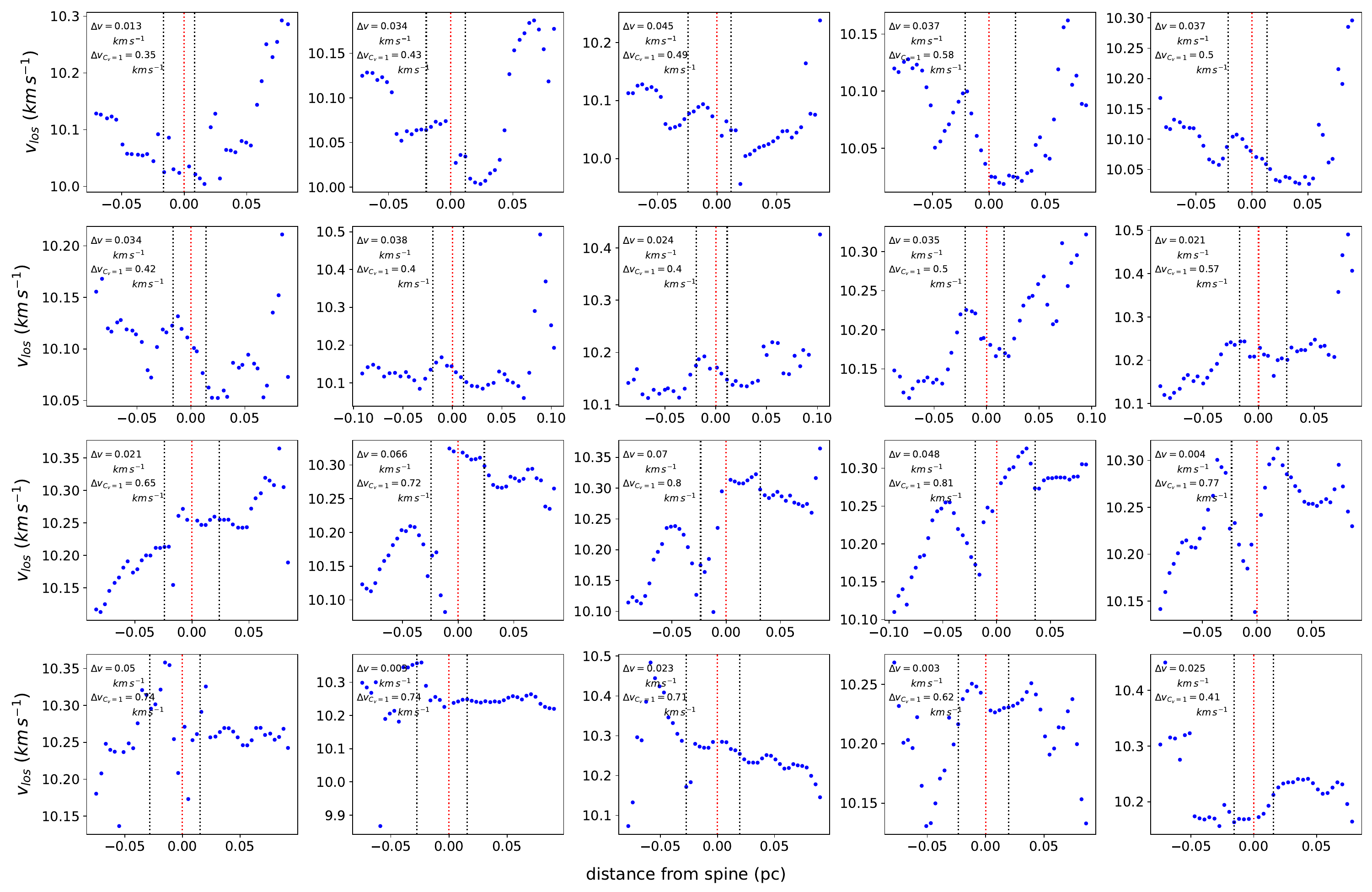}
  \caption{Line of sight velocity profiles in the orthogonal cuts in filament-2. The red-dotted vertical lines show the position of the filament spine and the black-dotted vertical lines show the boundary of the filament for the respective orthogonal cuts shown. 
  The $\Delta {\rm v}$ parameter ($\Delta {\rm v}_h$ in Eq. \ref{eq_cv-chen}) for the cut is shown in the upper-left corner. The $\Delta {\rm v}_h$ value needed to have $C_{\rm v} = 1$ ($\Delta {\rm v_{C_V=1}}$) is also shown.
  }
     \label{fig_vel_grad_cv_fil2} 
     
\end{figure}

Figs. \ref{fig_vel_grad_cv_fil1} and \ref{fig_vel_grad_cv_fil2} show the gradient in velocity across cuts orthogonal to the major axes of the filaments 1 and 2, respectively. These cuts are spaced 1 beam apart, and are perpendicular to the filament spine at that location.

\end{multicols}

\FloatBarrier

\FloatBarrier

\begin{multicols} {2}

Figure \ref{fig_kde_cv} shows the Kernel Density Estimate (KDE) distribution of \cv values across the two filaments. This distribution highlights that the values have a strong distribution peak below \cv = 0.01, and a smaller secondary peak at $\approx 0.035$.

\begin{Figure}  
\centering
\includegraphics[width=0.97\textwidth]{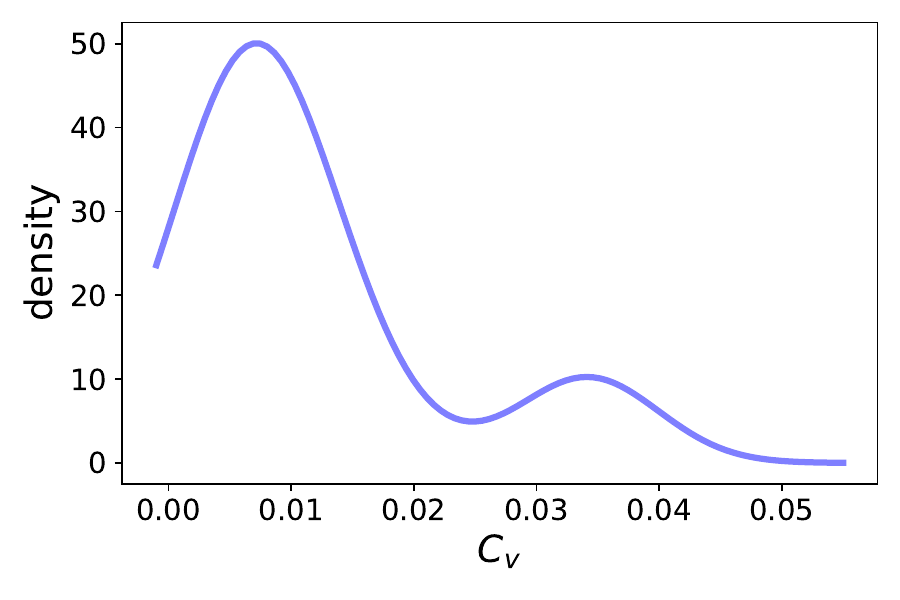}
  \captionof{figure}{Kernel density estimate of the distribution of the \cv values across both filaments.}
     \label{fig_kde_cv} 
     
\end{Figure}
\FloatBarrier

\section{Comparison with filament formation scenarios considered in \citet{chen_2020_fila_form}}
\label{app_chen_compare}

As mentioned in Sect. \ref{sec_fila_form}, there is a systematic and significant difference between the \cv values presented in this work (\cv $\ll 1$) and those from \citet{chen_2020_fila_form} simulations (\cv $\sim 0.3$).

One key difference between the two works is in the mass-per-unit length ($M/L$) of the filaments. \citet{chen_2020_fila_form} considered $M/L \sim$ 10 \msun\ pc$^{-1}$ in their simulations, whereas for the filaments in B5, $M/L \approx$ 80\msun\ pc$^{-1}$ \citep{anika_2021}.
The $C_{\rm v} \ll 1$ values for the B5 filaments could also imply that the filament formation in B5 was rather cylindrically symmetric \citep[closer to the scenario depicted in the right part of Fig. 1 in][]{chen_2020_fila_form}, as opposed to being formed in a planar geometry, which was explored by their simulations (left part in the same figure).

Another possible difference could be the magnetic fields present in the filaments. As mentioned in Sect. \ref{sec_fila_form}, for the required support against gravity in the B5 filaments, the estimated magnetic field is poloidal and parallel to the spine of the filament, with a strength of $\sim 500\, \rm\mu$G.
This would suggest that the strength and structure of the magnetic field is vastly different from that in the simulations of \citet{chen_2020_fila_form}, where the magnetic fields are $\sim 10\, \rm\mu$G, and roughly perpendicular to the filament spine.

While some velocity profiles in the orthogonal cuts to the filaments (Figs. \ref{fig_vel_grad_cv_fil_exmpl}, \ref{fig_vel_grad_cv_fil1}, and \ref{fig_vel_grad_cv_fil2}) have similar shapes to those in simulations of \citet{chen_2020_fila_form},  the absolute difference in velocities at the boundary in these cuts is still relatively small.
Only 6 out of the 34 orthogonal cuts show a velocity gradient $\ge0.1\,\kms$ at the filament boundary (black-dotted lines in Figs. \ref{fig_vel_grad_cv_fil_exmpl}, \ref{fig_vel_grad_cv_fil1}, and \ref{fig_vel_grad_cv_fil2}), and none of them exceed $0.15\,\kms$. 
In comparison, the velocity gradient seen in the filaments in the simulations of \citet{chen_2020_fila_form} is $\gtrsim0.15\,\kms$, except in the early stages. This could be another potential factor behind the difference in the \cv values between the two works.

\end{multicols}

\end{appendix}

\end{document}